%% file: main.tex
\documentclass[aps,pra,a4paper,twocolumn,longbibliography,groupedaddress]{revtex4-2}

\usepackage{amsmath}
\usepackage{algorithm}
\usepackage{algpseudocode}
\usepackage{enumitem}
\usepackage{amssymb}
\usepackage{amsfonts}
\usepackage{braket}
\usepackage{dsfont}
\usepackage{MnSymbol} 

\usepackage{color}
\usepackage[dvipsnames]{xcolor}
\usepackage{amsmath}
\usepackage{tikz}
\usepackage{qcircuit}
\usepackage{bm}

\usepackage{graphicx, import}
\usepackage{soul}
\usepackage[utf8]{inputenc}
\usepackage[export]{adjustbox}
\usepackage{parskip} 

\usepackage{hyperref}

\hypersetup{
    colorlinks,
    citecolor={blue},
    urlcolor={blue}
}

\usepackage[nonumberlist, acronym, toc]{glossaries}

\makeatletter
\newcommand*{\glsplainhyperlink}[2]{%
    \begingroup%
      \hypersetup{hidelinks}%
      \hyperlink{#1}{#2}%
    \endgroup%
}
\let\@glslink\glsplainhyperlink
\makeatother

\usepackage[normalem]{ulem}

\begin{document}
\include{glossary}

\title{
Potential of quantum scientific machine learning \\ applied to weather modelling}

\author{Ben Jaderberg}
\author{Antonio A. Gentile}
\author{Atiyo Ghosh}
\author{Vincent E. Elfving}
\affiliation{PASQAL, 7 rue Léonard de Vinci, 91300 Massy, France}

\author{Caitlin Jones}
\author{Davide Vodola}
\affiliation{BASF Digital solutions, Next Generation Computing, Pfalzgrafenstr.~1, 67061 Ludwigshafen, Germany}
\author{John Manobianco}
\affiliation{BASF Corporation, APM/DS, 2 TW Alexander Drive, Research Triangle Park, North Carolina 27709, USA}
\author{Horst Weiss}
\affiliation{BASF SE, Next Generation Computing, Pfalzgrafenstr.~1, 67061 Ludwigshafen, Germany}


\begin{abstract}

In this work we explore how quantum scientific machine learning can be used to tackle the challenge of weather modelling. Using parameterised quantum circuits as machine learning models, we consider two paradigms: supervised learning from weather data and physics-informed solving of the underlying equations of atmospheric dynamics. In the first case, we demonstrate how a quantum model can be trained to accurately reproduce real-world global stream function dynamics at a resolution of 4°. We detail a number of problem-specific classical and quantum architecture choices used to achieve this result. Subsequently, we introduce the barotropic vorticity equation (BVE) as our model of the atmosphere, which is a 3\textsuperscript{rd} order partial differential equation (PDE) in its stream function formulation. Using the differentiable quantum circuits algorithm, we successfully solve the BVE under appropriate boundary conditions and use the trained model to predict unseen future dynamics to high accuracy given an artificial initial weather state. 
{Whilst challenges remain, our results mark an advancement in terms of the complexity of PDEs solved with quantum scientific machine learning.
}
\end{abstract}

\maketitle


\section{Introduction} \label{sec:introduction} 

The ability to forecast the weather is of great importance to many areas of modern society including agriculture, transportation and the mitigation of extreme weather events. Accurate prediction of near and medium term weather remains difficult owing to the complex, non-linear and dynamic behaviour of atmospheric physics. Simulating such processes incurs a staggering computational cost. For example, the \gls{ifs} used by the \gls{ecmwf} computes numerical simulations on a 1-million CPU core supercomputer every 6 hours~\cite{ecmwfFactSheet, ecmwfMediumrangeForecasts}. 
{These state-of-the-art \glspl{sem} involve computations that alternate between real-space and spectral representations at every time step~\cite{orszag1970transform, eliasen1970numerical}, as discretised in a semi-Lagrangian scheme~\cite{hortal2002development} where the grid computation often relies upon a finite element method~\cite{vivoda2018finite}. However, significant open problems remain with these methods, including maintaining an appropriate grid resolution, avoiding numerical instabilities and bottlenecks imposed by data transfer when transposing from real to spectral spaces~\cite{wedi2014increasing}. This ultimately leads to error in predictions~\cite{grams2018atmospheric}, particularly at time scales beyond 1 day~\cite{lavers2019forecast, han2021deep}. Crucially, it is not clear yet how much further these numerical methods will need to scale to solve these issues, nor if the required computational power to do so will be available. Therefore, methods which might offer reductions in computational cost or improved long-time accuracy are highly desirable.}

Recently, \gls{ml} has demonstrated great potential as a solution. \gls{ml} methods differ from traditional methods as they make predictions based on weather patterns that are present in the data sets used to train them. Rather than repeatedly running expensive simulations, a larger one-off cost is invested in the training of a \gls{ml} model, with the idea that after training the model can make predictions at low cost. When large amounts of historical data are used as a data set~\cite{nguyen2023climax}, the predictions of such models may even outperform \gls{ifs} in accuracy~\cite{bi2022pangu, chen2023fuxi}.
However, open questions remain around whether such models generalise to regions with limited data or under extreme conditions not seen previously~\cite{reichstein2019deep}. This is in contrast to some traditional numerical methods, which do not face this issue as they forecast purely based on the underlying \glspl{pde} of atmospheric simulations.
  
Therefore, it seems natural to try to combine these approaches. This has led to the field of \gls{sciml}, in which the \gls{ml} model aims to also satisfy the physical laws of nature. Promisingly, there is already evidence that for weather and climate modelling this paradigm leads to less data required for training and improved generalisation of the trained models~\cite{kashinath2021physics}. 
In particular, among the various \gls{sciml} approaches, the \gls{pinn} formulation has led to a rapid progression of impressive results~\cite{wu2020enforcing, manepalli2019emulating, beucler2019achieving, bihlo2022physics}. 
Here the output of a neural network is used as a solution to the underlying \glspl{pde}, achieved via introducing a penalty during training if the output at a discrete number of sampled locations does not satisfy the equations. Notably,
\glspl{pinn} have been shown to demonstrate various benefits, including downscaling (i.e., accurately predicting at a finer resolution than the training data) ~\cite{stengel2020adversarial} and improved extrapolation to unseen conditions~\cite{esmaeilzadeh2020meshfreeflownet}.

The success of classical \gls{ml} has led to the swift development of \gls{qml}. Very broadly \gls{qml} is any machine learning approach that uses a quantum computer for all or part of the process. The aim is to leverage the properties of a quantum computer to solve machine learning problems more efficiently, most ambitiously to tackle classically intractable problems.
Encouragingly, there are a number of \gls{qml} algorithms which demonstrate rigorous computational advantage, for example  with regards to specific learning problems~\cite{liu2021rigorous, lloyd2016quantum, huang2022quantum} or replacing fundamental sub-routines in classical \gls{ml}~\cite{harrow2009quantum, liu2024towards}.  However, due to the high number of quantum operations and qubits required, these specific algorithms require fault tolerant quantum computing, widely assumed to be on the order of a decade away.

An alternative approach, more amenable to the near term, is variational \gls{qml} algorithms~\cite{havlivcek2019supervised, schuld2019quantum, cerezo2021variational}. In this setting, a classical neural network is replaced with a quantum circuit in which the logic gates embed tunable parameters, leading to a tunable output. This output can then be used to calculate the value of a loss function and, following typical \gls{ml} procedures, can be used to optimise the weights to train the model. Importantly, variational \gls{qml} models require reduced gate depths and qubit counts~\cite{perdomo-ortiz2018} and exhibit intrinsic resilience to some noise sources, a desirable feature when using real quantum hardware as evidenced by many experimental demonstrations~\cite{hubregtsen2022training, abbas2021power, jaderberg2022quantum}.
Mirroring classical \gls{sciml}, variational \gls{qml} architectures can also be trained to satisfy a set of governing equations, leading to the field of \gls{qsciml}. Within this realm, a particular effort has been focused on using physics-informed terms analogous to \glspl{pinn}~\cite{kyriienko2021solving, paine2023physics, heim2021quantum}. 



In this work we use \gls{qsciml} to solve the \gls{bve}, a 3\textsuperscript{rd} order partial differential equation (PDE), for a global setting via a physics-informed variational quantum algorithm and ad-hoc circuit architectures. The corresponding model is then benchmarked aside its classical equivalent. 
This is strongly motivated by the fact that, as stochastic optimisation problems, the trainability and performance of \gls{qsciml} models are hard to predict in general. This makes it all the more important to test them in new domain areas, as is reflected by progress across the field in other application areas~\cite{ghosh2022harmonic, varsamopoulos2022quantum, kumar2022integral}.

This work is laid out as follows. In section \ref{sec:bve} we introduce the problem setting including the barotropic vorticity equation (\gls{bve}), a \gls{pde} which we use as our model of the atmosphere to solve. In section \ref{sec:pqcs} we introduce how quantum computing circuits with parameterised gates can be trained as \gls{ml} models, either through data only or with the inclusion of \gls{pde}s.
{In section \ref{sec:specific_model_features}, we give further details on the encoding of the input parameters in the quantum circuit and on the ansatz used for the optimization.}
In section \ref{sec:results} we present results achieved using each paradigm, looking at artificial weather systems and cases of real global weather. Finally, we discuss and summarise our results in section~\ref{sec:conclusion}. 

\section{The barotropic vorticity equation}\label{sec:bve}

The \gls{bve} describes the evolution of the vorticity, a measure of the rotation of air, for an inviscid and incompressible flow. The \gls{bve} can be used to model a simplified atmosphere in which winds are independent of geopotential height~\cite{shin2006critical, shaman2011atmospheric, tang2018nonlinear}. Considering the flow within a single 2D plane in this atmosphere, the \gls{bve} can be written as
\begin{equation}
    \frac{D\eta}{Dt} = \frac{\partial \eta}{\partial t} + (\vec{u} \cdot \nabla)\eta = 0,
    \label{eqn:bve_cartesian}
\end{equation}
where ${u}$ is the velocity, $t$ is time and the absolute vorticity $\eta = \zeta + f$ is a sum of the relative vorticity of the flow $\zeta$ and the vorticity caused by the Earth's rotation $f$.

It is convenient to transform Eq.~\eqref{eqn:bve_cartesian} into spherical coordinates by defining the longitude $\lambda$, the latitude $\phi$ and $f=2 \Omega \sin{\phi}$ where $\Omega = 7.292 \times 10^{-5}$ rad/s is the angular velocity of the Earth. It is also convenient to introduce the stream function $\psi$, linked to the relative vorticity via  $\zeta = \nabla^2 \psi$, for which lines of constant value are streamlines (i.e., curves which show the direction in which a fluid element will travel at any point in time). In terms of stream function and vorticity, Eq. (\ref{eqn:bve_cartesian}) can be written as
\begin{equation}
    \frac{\partial \zeta}{\partial t} = -\frac{2\Omega}{r^2}\frac{\partial \psi}{\partial \lambda} -  \frac{1}{r^2 \cos\phi}\left(\frac{\partial \psi}{\partial \lambda}\frac{\partial \zeta}{\partial \phi} - \frac{\partial \psi}{\partial \phi}\frac{\partial \zeta}{\partial \lambda} \right).
    \label{eqn:bve_spherical}
\end{equation}

Further still, one can express Eq. \eqref{eqn:bve_spherical} in terms of the streamline function alone
\begin{equation}
\small
       \begin{split}
       & F(\psi, \phi, \lambda, t)  = \\  & -\tan\phi\frac{\partial^2 \psi}{\partial \phi \partial t} + \frac{\partial^3 \psi}{\partial \phi^2 \partial t} + \frac{1}{\cos^2\phi}\frac{\partial^3 \psi}{\partial \lambda^2 \partial t} + 2\Omega \frac{\partial \psi}{\partial \lambda}  \\
      &+ \frac{1}{r^2 \cos\phi}\frac{\partial \psi}{\partial \lambda}\left( -\frac{1}{\cos^2\phi} \frac{\partial \psi}{\partial \phi} - \tan\phi\frac{\partial^2 \psi}{\partial \phi^2} + \frac{\partial^3 \psi}{\partial \phi^3}\right. \\
      & \left. + 2\tan\phi\frac{1}{\cos^2\phi}\frac{\partial^2 \psi}{\partial \lambda^2}  + \frac{1}{\cos^2\phi}\frac{\partial^3 \psi}{\partial \lambda^2 \partial \phi} \right) \\
      & - \frac{1}{r^2 \cos\phi}\frac{\partial \psi}{\partial \phi}\left(-\tan\phi\frac{\partial^2 \psi}{\partial \phi \partial \lambda} + \frac{\partial^3 \psi}{\partial \phi^2 \partial \lambda} + \frac{1}{\cos^2\phi}\frac{\partial^3 \psi}{\partial \lambda^3} \right) \\
      & = 0,
    \end{split}
\label{eqn:bve_as_streamline}
\end{equation}
where for conciseness in the following, in the last line we omitted the explicit dependencies of the functional $F$ on partial derivatives and of $\psi$ on the features. 
Further details of these derivations can be found in Appendix \ref{app:bve_derivation}. Eq. (\ref{eqn:bve_as_streamline}) represents the form of the BVE used as the \gls{pde} training loss in section \ref{subsec:dqc_results}.

\section{Quantum circuits as machine learning models}
\label{sec:pqcs}

\begin{figure*}
    \centering
    \includegraphics[width=.9\textwidth]{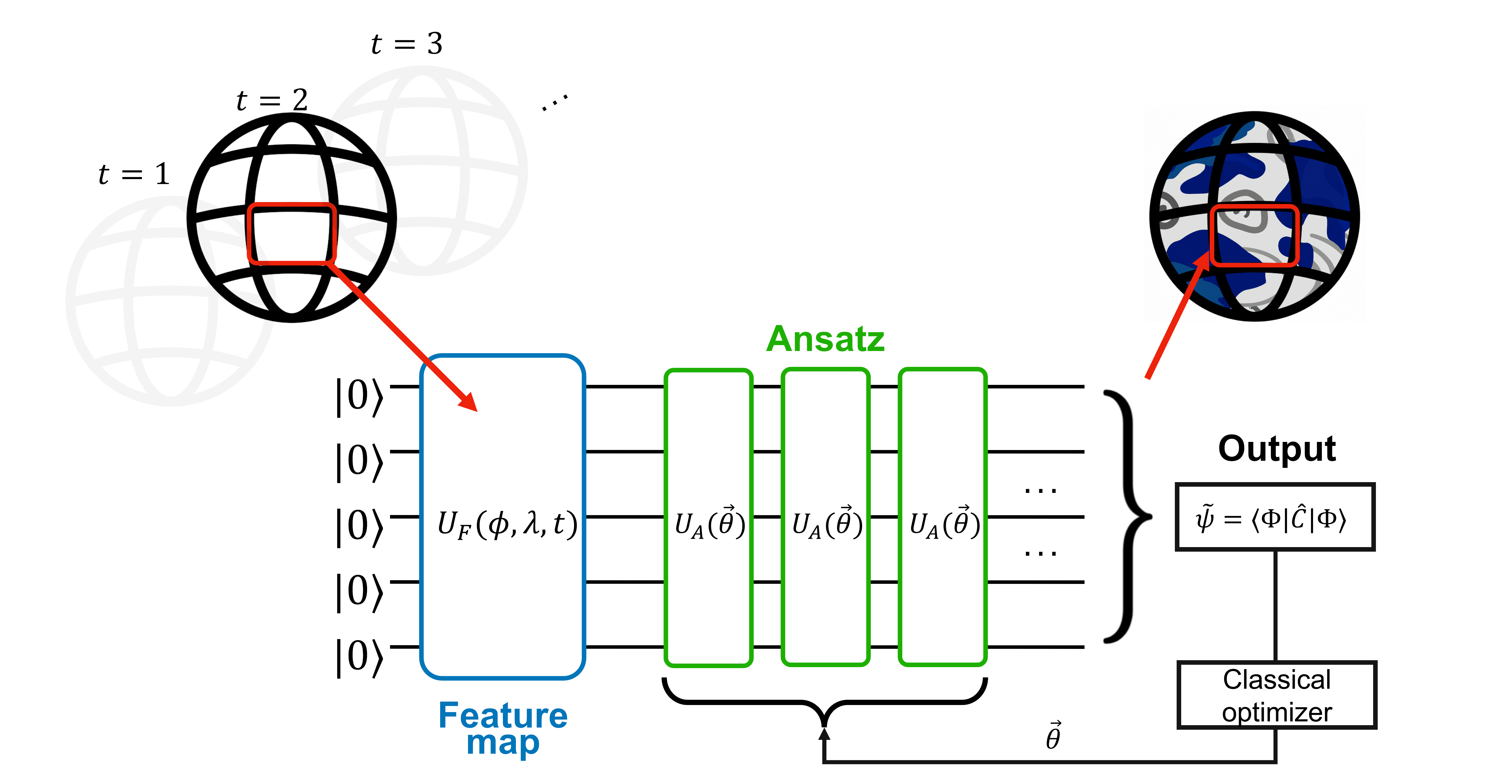}
    \caption{An overview of the \gls{qsciml} algorithm used in this work. Considering a single layer of the atmosphere, a collocation point can be defined in terms of the latitude $\phi$, longitude $\lambda$ and time $t$, which is embedded into the Hilbert space using a quantum feature map. Subsequently, an ansatz consisting of trainable unitary operations are applied to manipulate the embedded features. Finally, measurement of qubits are used to construct an expectation value of a cost operator $\hat{C}$, which is used as a prediction of the stream function $\tilde{\psi}$. The model is trained using the 
    \acrfull{dqc} algorithm to minimise the loss function associated to the \gls{bve} and the boundary conditions.}
    \label{fig:pqc_circuit_diagram}
\end{figure*}

A quantum algorithm, consisting of a set of quantum (parameterised) logic gates, can be visualised as a quantum circuit diagram. One example of this is shown in Fig.~\ref{fig:pqc_circuit_diagram}, which describes at a high-level the variational \gls{qml} algorithm used in this work \cite{cerezo2021variational}. The diagram should be read left-to-right, with each line representing a qubit, initialised as standard in the computation basis state $|0\rangle$ and evolved according to the various gates overlapping with the corresponding lines.
Looking at this diagram, the first component is the quantum \gls{fm}. This is used to encode a given classical input $\vec{x}$ into the Hilbert space, for which $\vec{x}$ is a (multi-dimensional) coordinate from the problem domain. In order to do so, after initialisation the qubits are acted upon by a unitary matrix $\hat{U}_F(\phi, \lambda, t)$. Lower-level details about the specific encoding unitary are given in section \ref{sec:specific_model_features}.


Following the \gls{fm} is a second component of our algorithm, the variational ansatz. Each layer of the ansatz can be expressed mathematically as a general unitary operation $\hat{U}_A(\vec{\theta})$ acting across all of the qubits and parameterised by the vector of tunable parameters $\vec{\theta}$. 
In practice, each block $\hat{U}_A$ is composed of a series of single-qubit operations and multi-qubit entangling operations and depends on a small subset of parameters $\Vec{\theta_i}$, such that $\hat{U}_A(\vec{\theta}) = \hat{U}_{A_1}(\theta_1) \hat{U}_{A_2}(\theta_2)\hat{U}_{A_3}(\theta_3)... $. 
Importantly, each parameter $\theta_i$ is not prescribed like those of $\hat{U}_F$, but instead is tuned during the learning by a classical optimiser. 
A particular case is given by digital-analog quantum architectures, where the unitaries acting upon (subsets of) the circuit qubits correspond to the spontaneous (driven) evolution over a time $t$ of the subtended quantum system $\hat{U}_{A_i} (\Vec{\theta_i}, t) = \exp{(-i \hat{H}_i(\vec{\theta}_i) t)} $, as described by the corresponding Hamiltonian $\hat{H}_i$.

Finally, with the input coordinate encoded by the \gls{fm} $\hat{U}_F(\phi, \lambda, t)$ as a quantum state, which is then transformed by the trainable ansatz $\hat{U}_A(\vec{\theta})$, classical information is extracted from the quantum model through a set of measurements of the evolved state $|\Phi\rangle = \hat{U}_A(\vec{\theta}) \hat{U}_F(\phi, \lambda, t) |0\rangle$. 
In \gls{qnn} architectures, these measurements are designed to obtain the expectation value $\langle\Phi|\hat{C}|\Phi\rangle$  with respect to a desired cost operator $\hat{C}$. Alternatively, the overlap $\braket{\Psi|\Phi}$ with another quantum state $\ket{\Psi}$ used as a reference is the approach adopted in ``kernelised'' versions. 
We choose the \gls{qnn} approach, for which the choice of cost operator is another crucial architectural choice~\cite{kyriienko2021solving}. In this work we adopt the cost operator $\hat{C} = \sum^N_{m=1} \hat{Z^m}$, an equally-weighted total magnetization across the $N$ qubits. The expectation value of this operator, extracted from the quantum circuit, is then taken as the prediction of the stream function $\psi$ for the given input coordinate. Importantly, the combination of constant depth ansatz and 1-local observables is known to avoid trainability issues~\cite{cerezo2021cost}. 

\subsection{PQC training routine on data}

In section \ref{sec:pqcs} we introduced \glspl{pqc} as the variational quantum architecture of choice, to effectively embed a number of real parameters into a quantum circuit and tune the dependency of the circuit output(s) on the corresponding input values. 

We are thus equipped to describe how to perform supervised learning on a training data set $\{\vec{x}_i, {y}_i \}$, i.e. train the quantum circuit to reproduce the dependency of the labels ${y}_i$ upon an (unknown) function $f$ acting on the model features, i.e.  ${y}_i = f (\vec{x}_i)$. 
Note that here we assume the output ${y}$ is a scalar for simplicity. Thus, the measured output of the \gls{pqc} 
\begin{equation}
\tilde{y}_i = \Braket{\Phi (\vec{x}_i ) | \hat{C} | \Phi (\vec{x}_i )}
\label{eq:qnn_output}
\end{equation}
can be used to to approximate the given $y_i \forall i$.
Optimising a model so that $\tilde{y}_i = y_i$ can be cast as a regression problem, as shown in~\cite{mitarai2018quantum}, by introducing an appropriate loss function $L$, capturing the distance from the provided training labels and the circuit outputs. This strategy was coined \gls{qcl}. A typical choice, adopted also in this paper, is a \gls{mse} error $L \equiv \sum_i (\tilde{y}_i - y_i)^2/N_I$ with $N_I \equiv |\{\Vec{x}_i\}|$ the number of training points; however other options are possible~\cite{mitarai2018quantum, kyriienko2021solving}. 
Finally, this can be coupled with (gradient-based) optimization methods and back-propagation, in order to progressively train the free variational parameters $\Vec{\theta}$ of the circuit, until $L$ is minimized upon convergence by iterating the procedure. 

\subsection{Solving nonlinear PDEs with quantum computers}\label{subsec:solving_pdes}

When targeting the solution of \gls{pde}s, several quantum computing strategies have been suggested. Often invoked for this purpose is the HHL algorithm, a linear algebra quantum subroutine with exponential improvements for some classes of problems~\cite{harrow2009quantum}. Whilst later HHL was followed by a number of improvements and extensions~\cite{lloyd2020quantum, duan2020surveyhhl}, strategies within this category are still either limited to address linear problems, rely upon linearization techniques with restrictive applicability, or require ancillary quantum registers to tackle increasing degrees of non-linearity in the problem.
On the contrary, we are often interested in solving non-linear, high-dimensional problems. The large overhead to solve these problems with traditional computing makes them ideal candidates to test the feasibility and potential advantage of (hybrid) quantum architectures. 

This premise has led to the development of \gls{nisq}-amenable \gls{qml} approaches. Among them, the \gls{dqc} algorithm optimises the variational parameters of the quantum circuit using a loss function informed by the (known) equations governing the problem \cite{kyriienko2021solving}. \gls{dqc} is a natural choice to explore with near-term quantum hardware implementations, due to its computationally cheap encoding and readout circuit components, whilst capable of addressing nonlinear instances of PDEs without the need for any preliminary linearization scheme.

\paragraph{DQC training routine.} 
In this paragraph, we detail the essential steps of \gls{dqc}, a more detailed explanation can be found in \cite{kyriienko2021solving}.
First, we obtain a trial solution by evaluating 
the output of a variational quantum circuit $\Tilde{y_i} (\Vec{x}_i, \Vec{\theta})$ as in Eq.~\eqref{eq:qnn_output}, which can approximate the action of any function $f(\Vec{x})$ on the feature(s) $\Vec{x}$ (here $\psi$ and $\Vec{x} = (\phi, \lambda, t)$, respectively), i.e. represent a tentative solution for the equations targeted. 
The main idea behind \gls{dqc} is to variationally train such trial solutions $\Tilde{y}$, by quantitatively evaluating their accuracy, until the training eventually converges to an approximate solution $\Tilde{y} \equiv \Tilde{f}(\Vec{x}) \sim f(\Vec{x})$.  
Inspired by \glspl{pinn}, such training can occur in a physics-informed manner by incorporating a loss function $\mathcal{L}_\text{PDE}$ derived from a \gls{pde}. In this work, that \gls{pde} is specifically the \gls{bve}, as formulated in Eq. (\ref{eqn:bve_as_streamline}). In order to express derivative terms in the equation under the same framework, it is also necessary to build and evaluate derivative quantum circuits to estimate $\partial \Tilde{f}(x) / \partial x$, etc.
Specifically for our case, evaluating Eq. (\ref{eqn:bve_as_streamline}) requires four 1\textsuperscript{st} order derivatives, four 2\textsuperscript{nd} order derivatives and six 3\textsuperscript{rd} order derivatives. 
A quantitative metric for the accuracy of tentative solution can be the \emph{interior loss function} $\mathcal{L_\text{PDE}}$, which is the residual \gls{lhs} of the \gls{bve}, when replacing the formal with the trial solution $\psi(\phi, \lambda, t) \rightarrow \tilde{\psi}(\phi, \lambda, t) $. The evaluation of $\mathcal{L_\text{PDE}}$ is conveniently done at a discrete number of locations in the domain $\{\Vec{x}_i\} \in X$, such that the solution is tested across the whole domain, rather than at a single point.
Other loss terms can be introduced, to quantify the adherence to boundary conditions, or take into account regularisation data points $\{y_i\}$ attained elsewhere. 
Combining all the available loss e.g. as a sum of all the independent loss terms, one constructs a total loss $\mathcal{L}$ \cite{raissi2017physics} as:
\begin{align}
    \label{eq:generic_loss}
    \mathcal{L} &= \mathcal{L}_{\text{PDE}} + \mathcal{L}_{\text{b.c.}} + \mathcal{L}_{\text{data}} = \nonumber \\
    &= \sum_{\{x_i\} \in X} L(F(\Tilde{\psi}, \Vec{x}_i)) +  \mathcal{L}_{\text{b.c.}} + 
    \sum_{\{y_i\}} L(y_i, \Tilde{\psi}(\Vec{x}_i)),
\end{align}
where $L$ is an appropriate distance function and $F$ follows the notation adopted in Eq.~\eqref{eqn:bve_as_streamline}.
A classical optimiser can then used to update the circuit weights $\Vec{\theta}$ in the \gls{dqc} in the direction which minimises the total loss function, e.g. via gradient descent \cite{kyriienko2021generalized}. 
With the parameters updated, this concludes one iteration of the \gls{dqc} algorithm, at which point a new and improved trial solution is generated and evaluated. This loop continues for a pre-determined number of epochs, or until a desired loss value is reached.

\section{Specific architecture choices}\label{sec:specific_model_features}

Whilst sections \ref{subsec:qcl_results} and \ref{subsec:dqc_results} give specifics of the quantum models used (e.g., number of qubits, ansatz layers, etc.), here we first detail some of the more advanced aspects used across all the numerical experiments in this work.
\\

\paragraph{Learnable linear feature transformation}
As discussed in section \ref{sec:pqcs}, the measured output of our quantum model is used as a prediction of the stream function $\psi$. Obtaining a close mapping between the model output and problem range is important for performance.

However, considering the specific chosen cost function $\hat{C} = \sum^N_{m=1} \hat{Z^m}$, any prediction of $\psi$ by the \gls{qnn} is bounded by the range of the sum of the $N$ single-qubit measurements, corresponding to a range of $[-N, N]$. By contrast, the real value of $\psi$ in the data ranges from $[-2, 2]$. 
One possible solution to this issue is to manually scale the output of the \gls{qnn} to be larger or smaller. However, this process can be tedious and inefficient. Instead, we apply a strategy that will work generally across different problem settings by introducing a learnable scaling and shifting to the output of the \gls{qnn}. Therefore, for a particular coordinate $\text{QNN}(\lambda, \phi, t)$, we apply the classical learnable linear transformation such that $\psi = \alpha_{n-1}\text{QNN}(\lambda, \phi, t) + \alpha_n$. 

Furthermore, given an input feature (e.g. $\phi$), we apply a similar transformation such that the model actually encodes $\alpha_1\phi + \alpha_2$ to conveniently span a range accessible to an angle encoding $[0, 2\pi)$. A similar process is also applied to $\lambda$ and $t$ with their own unique parameters. These additional trainable parameters are exposed to the optimiser alongside the parameters within the \gls{qnn}. 
Ultimately, we use learnable linear feature transformations to allow the model to automatically choose scales of input and output which are optimal for minimising the loss.
\\

\paragraph{Spherical geometry encoding}

In the problem under study, the input coordinates are $\vec{x} = \phi, \lambda, t$ as introduced in section \ref{sec:bve}. 
A vital component of our architecture involves consideration of the geometry of the problem. Specifically, for a problem domain over the surface of the Earth, approximated as a sphere, embedding symmetries of spherical coordinates leads to a more effective model. We implement such a scheme through a differentiable classical layer that pre-processes the features as:
\begin{align}
    &x = \: \sin\phi \: \cos\lambda\\
    &y = \: \sin\phi \: \sin\lambda\\
    &z = \: \cos\phi.
\end{align}
In other words, we encode the original two spherical dimensions $\phi$ and $\lambda$ into three spatial dimensions, which are in fact the Cartesian coordinate representations $x, y$ and $z$. Subsequently, the values of $t, x, y, z$ are embedded in the \gls{qnn} using the quantum \gls{fm}. We use the so-called angle encoding such that each dimension $ r \in \{t, x,y,z\}$ is encoded via the unitary $\hat{U}_F(r) = \bigotimes_{m}^N e^{-\frac{i}{2}\hat{Y}_m r}$, corresponding to a single-qubit Pauli $\hat{Y}$ rotation on each qubit.
\\

\begin{figure*}
    \centering
    \includegraphics[width=\textwidth]{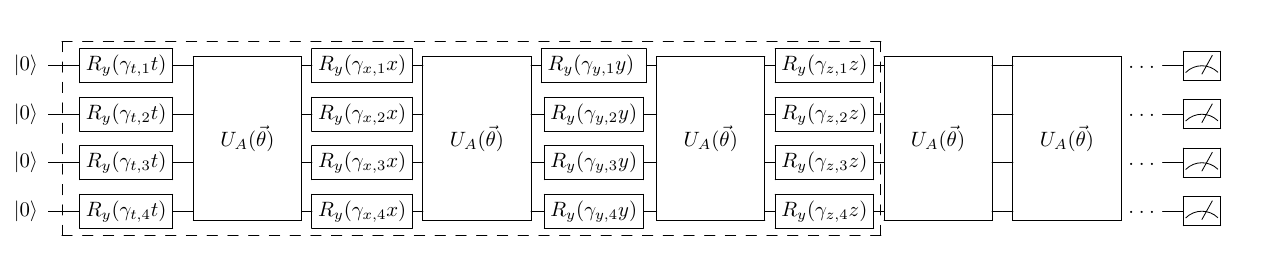}
    \caption{Circuit diagram demonstrating the specific quantum feature map used for \gls{dqc} experiments in this work. The original features $t$, $\phi$, $\lambda$ are mapped to $t$, $x$, $y$, $z$ through a classical pre-processing, which are then embedded in the quantum circuit by single-qubit $R_y$ rotation gates. Each feature is encoded sequentially in what is known as a serial quantum feature map, with trainable ansatz layers between the features to ensure they do not sum together trivially. Furthermore, for a given feature, each repeated encoding gate has its own independent trainable parameter, creating a \acrfull{tffm}~\cite{jaderberg2023let}. After the feature map {(enclosed in dashed line)}, the quantum model continues with trainable ansatz layers and eventually measurement.}
    \label{fig:qnn_architecture}
\end{figure*}

\paragraph{Serial quantum feature map}
It is possible to encode multi-dimensional coordinates into quantum models in two different ways; parallel and serial~\cite{schuld2021effect}. In this work we choose serial \gls{fm}s, in which each of the $d$ features are encoded across all $N$ qubits sequentially, for a total of $d$ encoding layers (see Fig.~\ref{fig:qnn_architecture}). 
This contrasts with a more direct \emph{parallel encoding}, whereby different sub-registers in the quantum circuit across $N_1, N_2, \ldots N_d$ qubits encode separately the various model features, such that $N_1 + N_2 + \ldots N_d = N$. 
Note that for a serial \gls{fm}, some intermediate gates are required between each feature to change the computational basis, otherwise the rotations of each feature will be summed leading to a loss of information. In this paper we choose these intermediate layers to be the same structure as our trainable ansatz.
{Repeated blocks of this \gls{fm} are also possible, to re-upload the features similarly to \cite{perez2020data}}.

The benefit of the serial \gls{fm} is apparent when interpreting the quantum model's expressive power in terms of a spectral decomposition into Fourier-like basis functions~\cite{schuld2021effect}.
Specifically, the number of basis functions per feature is determined by the eigenvalues of the encoding operation. This is a monotonous function of the Hilbert space dimension spanned by the encoding register(s). Thus, by encoding each feature sequentially over the whole $N$ register, rather than in parallel over size $N/d$ registers, the quantum model has access to a richer basis set.\\

\paragraph{Trainable frequency feature map}
As shown in Fig. \ref{fig:qnn_architecture}, we use the angle encoding quantum feature map, such that for any given feature $r$: $\hat{U}_F(r) = \bigotimes_{m}^N e^{-\frac{i}{2}\hat{Y}_m r}$. However, it was recently demonstrated that including trainable parameters into the \gls{fm} generator can lead to practical performance improvements, including specifically for solving \gls{pde}s~\cite{jaderberg2023let}. We adopt usage of these \glspl{tffm} in this work by modifying the previously introduced \gls{fm} to be
\begin{equation}
\hat{U}_F(r, \vec{\gamma}) = \bigotimes_{m}^N e^{-\frac{i}{2}\gamma_{r,m}\hat{Y}_m r},
\end{equation}
where the parameters $\gamma_{r,m}$ are trained by the optimizer.

\paragraph{Digital vs (Digital-)Analog Ansatz}
{
In this work, our choice of ansatz ($U_A(\vec{\theta})$ in Fig.~\ref{fig:qnn_architecture}) is the \gls{hea}~\cite{kandala2017hardware}, a structure widely adopted across variational quantum algorithms due to its amenability to near-term quantum hardware. Underpinning this is the \gls{hea}s constant depth scaling with number of qubits, use of commonly available hardware connectivity and well-understood benefits even in shallow regimes~\cite{leone2022practicalHEA}. 
Perhaps most importantly, also the use of entangling gates native to the quantum hardware of choice can crucially affect the ansatz performance. For example, the cascaded CNOT entangling gates originally proposed for \gls{hea} require complex operations to be implemented in emerging neutral atom quantum processors~\cite{bluvstein2023logical}. 
In such cases, simpler hardware implementations of all models trained in this paper could be achieved by applying emerging digital-analog approaches~\cite{parra2020digital, michel2023blueprint}, where e.g. chained CNOT gates are replaced by global analog entangling operations~\cite{digitalanalog2022}. 
} 
These considerations are important for any future implementations of this approach on realistic quantum hardware, without affecting the noiseless simulations performed here. 

\section{Results}\label{sec:results}

\subsection{Data-based training}
\label{subsec:qcl_results}

\begin{figure*}
    \centering
    \includegraphics[width=\textwidth]{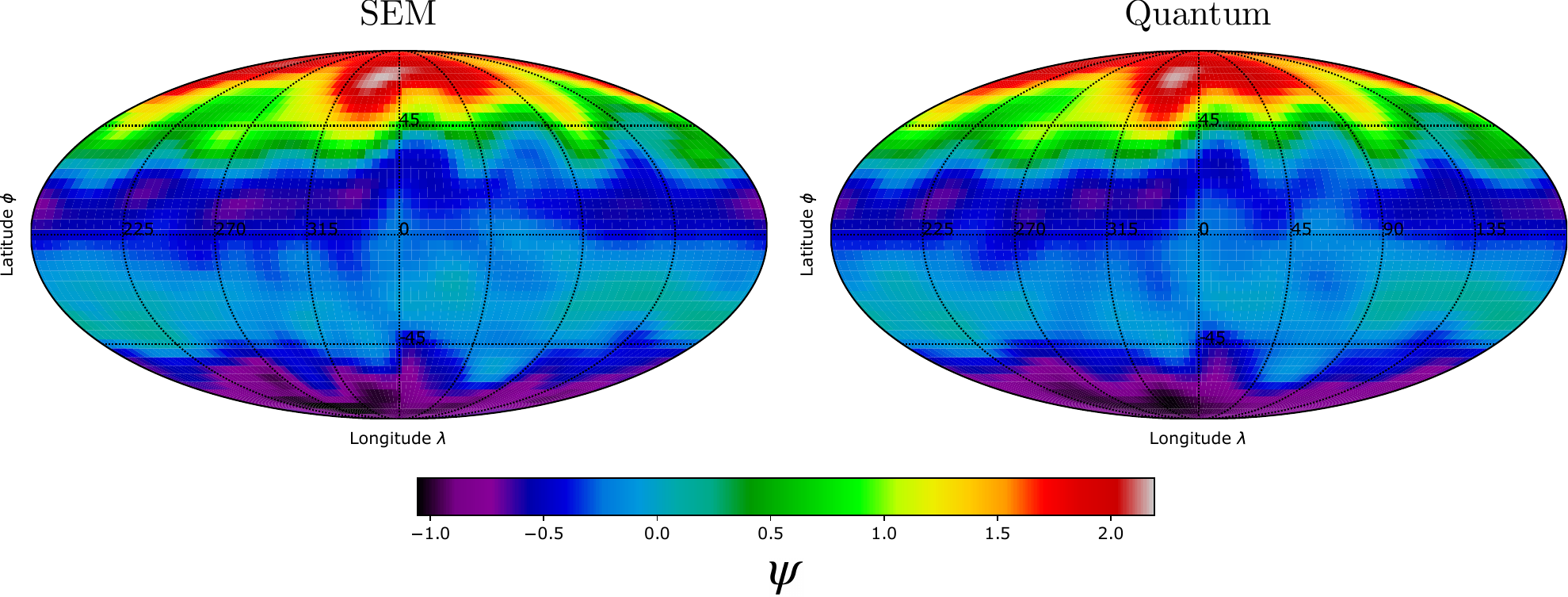}
    \caption{Prediction of global stream function $\psi$ after $t=22$ hours of evolution by \gls{sem} (left) or trained quantum model (right). The initial state is real weather data from midnight on 15th July 1980~\cite{hersbach2018era5}, down-sampled to a spatial resolution of $4.04^{\circ}$ (see Appendix \ref{app:real_world_data}). An \gls{sem} solver (see Appendix \ref{app:spectral}) is used as both a reference solution and to generate a data set from which the quantum model is trained.}
    \label{fig:qcl_result}
\end{figure*}

Having introduced the algorithm and architecture, here we present our results. In this first section we demonstrate the use of \gls{qsciml} for learning from data only, i.e. in a supervised learning setting. Here the learning problem is as such: given $\psi$ data at 3-hourly intervals $t=1, 4, 7,\dots, 22$, train the quantum model such that its output $\tilde{\psi}$ successfully reproduces $\psi$. In this experiment the data set used is derived from the ERA5 hourly relative vorticity on 15th July 1980~\cite{hersbach2018era5}, more information is given in Appendix \ref{app:real_world_data}.

In this experiment we train a $N=6$, $\ell=32$ layer \gls{qnn} with a \gls{hea} ansatz and serial \gls{tffm}. The \gls{hea} ansatz has $3N$ parameters per layer and the serial feature map contains $13N$ parameters in total, $9N$ from the three ansatz layers and $4N$ from the trainable encodings. Thus, our model has a total of $P=N\times(3\ell+13) = 654$ parameters. The model is trained using the Adam optimiser~\cite{kingma2014adam} for $N_{\text{iter}} = 5{,}000$ iterations with a batch size $b_s = 1602$ sampled randomly from the data and learning rate $lr = 10^{-2}$. The loss function is simply the \gls{mse} of the model prediction compared to the reference data $\mathcal{L} =  \mathrm{MSE}(\tilde{\psi}, \psi)$.

Once training is complete, we record the predicted stream function for each  point in space and time. To test the ability of the quantum model to reproduce the complex real-world weather patterns, we compare this prediction to the results from a standard approach for solving this equation, the spectral element method, see appendix \ref{app:spectral}, for two key \gls{fom}. These are the mean-relative error relative to the SEM median (MRE) and Pearson product-moment correlation coefficient (PPMCC), defined fully in Appendix \ref{app:fom}. Furthermore, given the MRE is a pointwise metric, we compute the median MRE at each time step. Similarly for the PPMCC, we define the median PPMCC as a single valued representation of how well correlated the predicted and reference dynamics are.

Overall we find excellent numerical agreement to the reference solution, showing that the model can successfully reproduce the stream function at space time points it was trained with. Across all 8 time steps, the model achieves between 7.1\% and 10.9\% median MRE and a median PPMCC of 0.870. The high correlation demonstrates how the quantum model correctly captures the different nontrivial dynamics across the globe, which qualitatively include an anticlockwise rotation around the north pole, a westwards drift around the equator and many other smaller patterns. Figure~\ref{fig:qcl_result} visualizes a single snapshot of this evolution at time $t=22$. Notably, the prediction of the quantum model is at a minimum resolution of $4.21^\circ$, comparable to that used in state-of-the-art \gls{ml} climate models~\cite{nguyen2023climax}.

\subsection{DQC: training on data and physics}
\label{subsec:dqc_results}

We now move to the more advanced task of predicting weather patterns by training the model to solve the \gls{bve}. This is achieved via the \gls{dqc} algorithm~\cite{kyriienko2021solving}, using boundary and initial conditions from the data set.
Specifically, the learning problem is posed as follows: given an initial stream function $\psi$ and   vorticity $\zeta$ and boundary conditions, train the quantum model to learn a stream function $\tilde{\psi}$ which solves the BVE, such that it successfully predicts unseen future $\psi$ and $\zeta$. Due to the additional complexity of solving an underlying \gls{pde}, as opposed to supervised learning, here we work on a lower-resolution artificial data set. 
This initial state is defined by a grid of 14 latitude ($\phi$) points and 25 longitude ($\lambda$) points over the entire sphere $\lambda \in (0^\circ, 360^\circ), \phi \in (-90^\circ, 90^\circ)$. Further description of the data is given in Appendix \ref{app:artificial_data}. 

The quantum model used is a $N=4$, $\ell=4$ \gls{qnn} with a \gls{hea} ansatz and serial \gls{tffm}, containing a total of 100 trainable parameters. 
The model is trained using the Adam optimiser for $N_{\text{iter}} = 30{,}000$ iterations with learning rate $lr = 10^{-2}$. In this experiment there are multiple loss terms
\begin{equation}
    \mathcal{L} = \alpha_1 \mathcal{L}_1 + \alpha_2 \mathcal{L}_2, + \alpha_3 \mathcal{L}_3, + \alpha_4 \mathcal{L}_4,
\end{equation}
including data terms $\mathcal{L}_1 = \mathrm{MSE}(\tilde{\psi}_{t=0}, {\psi}_{t=0})$ the \gls{mse} of the initial stream function, $\mathcal{L}_2 =  \mathrm{MSE}(\tilde{\zeta}_{t=0}, \zeta_{t=0})$ the \gls{mse} of the initial vorticity and $\mathcal{L}_3 =  \mathrm{MSE}(\psi_{t>0,\mathrm{equator}})$ the \gls{mse} of the stream function at the equator for times   $t \in (0, 3)$ at intervals of $0.1$. The inclusion of the $\mathcal{L}_3$ loss is necessary due to the fact that solving the BVE gives a unique $\zeta$ solution but a range of $\psi$ solutions up to a factor of integration, as seen from the relation $\zeta = \nabla^2 \psi$. Most notable is the loss $\mathcal{L}_4=\mathrm{MSE}(\mathrm{BVE})$, the \gls{pde} loss which is the \gls{mse} of the BVE residual (i.e., left-hand side of Eq. (\ref{eqn:bve_as_streamline})).

For the data loss terms, the loss weighting factors ($\alpha_1$, $\alpha_2$, $\alpha_3$) are the inverse of the mean of the respective training data squared (e.g, $\alpha_1 = 1/{|\bar{\psi}_{t=0}\vphantom{\big|}|^2}$). This ensures that data with larger scalar values are not disproportionately prioritised. The loss weight $\alpha_4=0.1$ was optimised via trial and error to minimise the corresponding total loss upon convergence. The batch size of each of the loss terms are $350$, $300$, $25$ and $350$ respectively, with sampled points (``collocation points'') chosen randomly from the data grid for the data loss terms, see caption of Fig.~\ref{fig:pqc_circuit_diagram}. For $\mathcal{L}_4$ there is no restriction that the points must lie within the data grid as Eq.~\eqref{eqn:bve_as_streamline} applies to all points so points are selected from the continuous problem space.

\begin{figure*}
    \includegraphics[width=\textwidth]{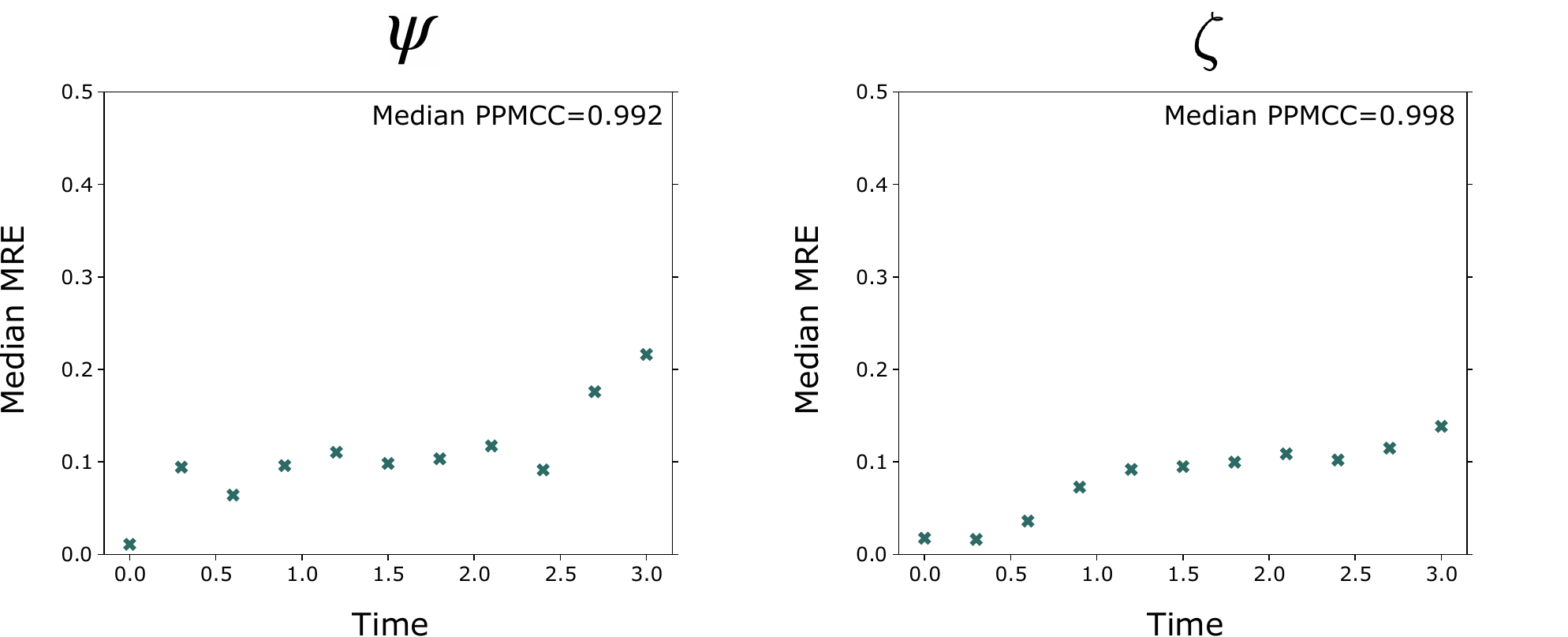}
    \caption{Performance of the \gls{dqc} model when predicting the stream function $\psi$ and vorticity $\zeta$ from an initial state by solving the \gls{bve}. The crossed markers indicate the spatially-averaged mean relative error (MRE) at each time point.}
    \label{fig:q6_final_FOM}
\end{figure*}

\begin{figure*}
    \centering
    \includegraphics[width=\textwidth]{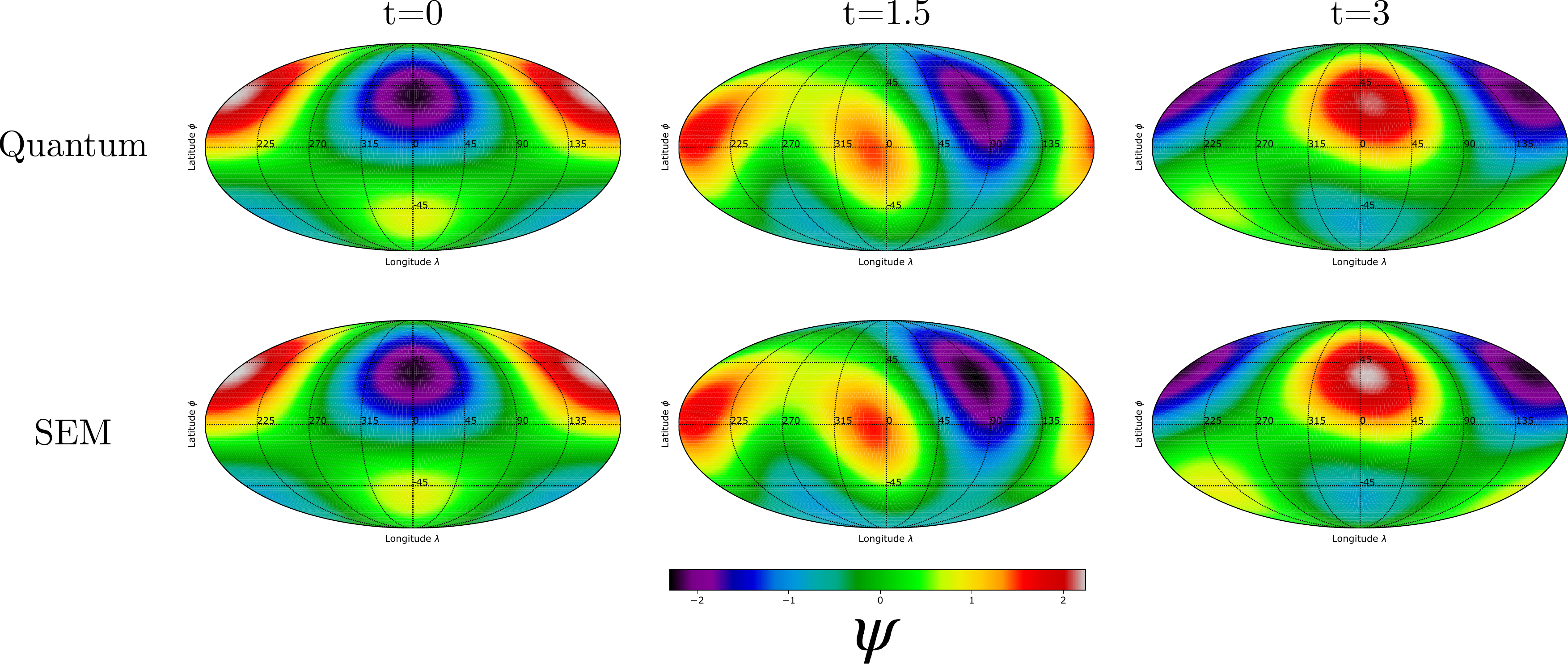}
    \caption{Visualisation of predicted stream function $\psi$ at times $t=0$, $t=1.5$ and $t=3$ by the trained quantum model and the \gls{sem} reference solution. The quantum model has access to the initial state, after which the evolution is predicted by solving the \gls{bve}. 
    }
    \label{fig:dqc_result}
\end{figure*}

Once training is complete, the model is used to predict the stream function and vorticity including at unseen time points $t>0$. Figure~\ref{fig:dqc_result} shows the quality of the solution for each observable. In terms of $\psi$ prediction, the model achieves a median MRE (left axis, crossed markers) of between 1.1\% at $t=0$ and 21.6\% at $t=3$. The predicted dynamics also have a median PPMCC (right axis, dashed line) of 0.994, very close to the maximally correlated value of 1. Similarly for $\zeta$, we see a median MRE of between 1.6\% and 13.8\% and a PPMCC of 0.998. Thus, for both figures of merit, the trained solution scores excellently. This demonstrates how a variational quantum model can solve the BVE. Furthermore, Figure~\ref{fig:dqc_result} visualises the predicted stream function at $t=0$, $t=1.5$ and $t=3.0$. Here we see the high level of similarity between the two solutions, as the northern negative vorticity rings rotate eastwards whilst the southern positive vorticity rings pushes northwards. Whilst at $t=0$ the solutions are identical, we do observe a small discrepancy in the grey region approximately located at $\lambda=0^{\circ}$, $\phi=30^{\circ}$ by $t=3$, reflecting what is shown by the figures of merit.

\section{Conclusion and Outlook}\label{sec:conclusion}

In this work we take the first steps in understanding how the field of \gls{qsciml} could be applied to weather modelling.
{In section \ref{sec:bve} we introduce the \gls{bve}, which we adopt as a simplified model of the global atmosphere for demonstrative purposes.}
Following this, in section \ref{sec:pqcs} we provide an overview of the general structure of variational quantum models, including a description of the data-based \gls{qcl} and \gls{dqc} learning algorithms which mirror the classical SL and \gls{pinn} approaches.  
Subsequently, in section \ref{sec:specific_model_features} we detail specific quantum architecture choices necessary to achieve high-accuracy results, including a geometric encoding for inputs on the surface of a sphere. Finally, in section \ref{sec:results} we demonstrate the use of the \gls{qcl} algorithm to predict the streamfunction across the spatial domain for times in the prediction range, as well as solve the BVE for the flow function in the case of \gls{dqc},  when a reference at a limited subset of points to ensure uniqueness is provided. 

From this trained model, we are able to successfully predict the evolution of an initial artificial weather state with high numerical accuracy for both the stream function and vorticity components. Specifically for the former one, we attain median $7.1\% \leq \text{MRE} \leq 10.9\%$ for the data-based model of $\psi$ in section \ref{subsec:qcl_results}, and $1.1\% \leq \text{MRE} \leq 21.6\%$ for the physics informed model of the same in section \ref{subsec:dqc_results}.

Our demonstration of solving the BVE is of particular note in the context of the wider literature. Specifically, in addition to being a three-dimensional problem, it is the highest order non-linear PDE to be tackled with the \gls{dqc} algorithm to date. {Even though no conclusive quantum advantage can be inferred from the results presented here, they yet demonstrate how \gls{qsciml} is steadily advancing to bigger and more complex problems.} 

It is clear of course that there is still a large gap to state-of-the-art classical ML solutions, which have been developed over the last few decades. Thus, it is important to consider what natural next steps arise from our work, and we do so distinguishing the cases of our supervised \gls{qcl} experiments and our physics-informed \gls{dqc} experiments.
For \gls{qcl}, in this work we evaluate the ability of quantum models to reproduce the training data, a necessary precondition to predicting unseen weather. With our positive results, looking forward a key goal is to numerically establish the generalisation capabilities of our model. The ability of a model to forecast beyond the training data is the basis for the adoption of machine learning, and understanding how specifically \gls{qsciml} models generalise is an active area of research~\cite{caro2022generalization, peters2023generalization}. Encouragingly, in section \ref{subsec:qcl_results} we already identified a temporal and spatial resolution which is both amenable to near-term quantum models and in which the predicted evolution of the global stream function contains meaningful dynamics. 

In terms of \gls{dqc}, a clear future target would be to extend our experiments to more realistic settings. 
Firstly, an initial state more representative of global weather patterns could be chosen, e.g. derived from real data itself. Secondly, one could drop the barotropic and incompressible assumptions behind the \gls{bve}. The ultimate goal here would be to solve for the hydrostatic primitive equations~\cite{iskandarani2003three}, employed by the state-of-the-art \gls{ifs} system~\cite{ecmwfNonhydrostaticFinitevolume}, and recently explored also in classical PINN settings ~\cite{wong2022learning, hu2023higher}. This more complex set of equations considers conservation of mass, momentum and energy and can also be coupled to moisture equations~\cite{taylor2011conservation}. 

Advancing our method to more complex and realistic models
will be more challenging due to 
(i) the linear scaling in complexity with the number of of training points, a problem affecting classical ML and numerical methods alike, 
along with (ii) the complexity in calculating high-order derivatives likely to appear in challenging PDEs, especially when performing classical simulations. 
Excitingly, recent advances in understanding how to evaluate \gls{dqc} in a mapped feature space~\cite{paine2023physics} raise the prospect of a quantum approach whose complexity scales as $\mathcal{O}(1)$ with the number of collocation points. Thus, understanding how to apply these new advances to the differential equations underpinning weather modelling would be an exciting and valuable next step on the route to advantageously exploiting the Hilbert space accessible to quantum devices.

\section*{Data Availability}
The ERA5 weather data~\cite{hersbach2018era5} was downloaded from the Copernicus Climate Change Service (C3S) (2023).


\bibliography{bibliography.bib}

\appendix
\newpage

\section*{Appendix}

\section{Derivation of \gls{bve} in terms of stream function}\label{app:bve_derivation}

Let us start with the general Navier-Stokes equation for a viscous incompressible flow for the velocity vector and derive the vorticity equation. From this, we derive the Barotropic eqations for a non-viscous, incompressible flow in two dimensions. 

The Navier-Stokes equation for a viscous incompressible flow is
\begin{equation}
    \frac{\partial \vec{u}}{\partial t} + \vec{u} \cdot \nabla \vec{u} = - \frac{1}{\rho}\nabla p + \vec{g} + \nu \nabla^2\vec{u},
\end{equation}
where $\vec{u}$ is the velocity vector, $p$ is pressure, $\rho$ is density and $\nu$ is the viscosity term and $g$ is the gravity term. 

Taking the curl of the above equation results in
\begin{equation}
    \nabla \times \frac{\partial \vec{u}}{\partial t} + \nabla \times (\vec{u} \cdot \nabla \vec{u}) = -\nabla \times (\frac{1}{\rho}\nabla p) + \nabla \times \vec{g} + \nabla \times (\nu\nabla^2 \vec{u}).
\end{equation}

Let us first look at the terms in the \gls{lhs}. From the relation $\vec{\omega} = \nabla \times \vec{u}$, the first term gives us $\nabla \times \frac{\partial \vec{u}}{\partial t} \equiv \frac{\partial \vec{\omega}}{\partial t}$. Furthermore, the term $\vec{u} \cdot \nabla \vec{u}$ can be written as,
\begin{equation}
    \vec{u} \cdot \nabla \vec{u} = \frac{1}{2} \nabla (\vec{u}\cdot \vec{u}) - \vec{u} \times (\nabla \times \vec{u}) = \nabla\frac{|\vec{u}|^2}{2} - \vec{u}\times \vec{\omega}
\end{equation}
Now taking the curl of this quantity results in,
\begin{equation}
\begin{split}
     \nabla \times (\vec{u} \cdot \nabla \vec{u}) &= \nabla \times \nabla\frac{|\vec{u}|^2}{2} - \nabla \times (\vec{u}\times \vec{\omega}) \\
     &= 0 + \nabla \times (\vec{\omega} \times \vec{u}) \\
     &= (\vec{u} \cdot \nabla)\vec{\omega} - (\vec{\omega} \cdot \nabla)\vec{u} + \vec{\omega}(\nabla \cdot \vec{u}) + \vec{u}(\nabla \cdot \vec{\omega})\\
     &= (\vec{u} \cdot \nabla)\vec{\omega} - (\vec{\omega} \cdot \nabla)\vec{u} + 0 + 0
\end{split}
\end{equation}
where we use the standard expansion $\nabla \times (\vec{A} \times \vec{B})$. We also use the fact that curl of a gradient is zero.  And in the last line, we notice that $\nabla \cdot \vec{u} = 0$ since the fluid is incompressible and $\nabla \cdot \vec{\omega} = \nabla \cdot (\nabla \times \vec{u}) = 0$ since divergence of a curl is zero.

For the RHS, we see that $\nabla \times (\frac{1}{\rho}\nabla p) = 0$ assuming the density is uniform and the fact that curl of a gradient is zero. Further, $\nabla \times \vec{g} = 0$ since $\vec{g}$ only has component in one direction (say the $\hat{z}$ direction). For the last term of RHS, we see that,
\begin{equation}
    \nabla \times (\nu \nabla^2\vec{u}) = \nu \nabla^2 \vec{\omega}
\end{equation}
Thus the Navier-Stokes vorticity equation looks like,
\begin{equation}
    \frac{\partial \vec{\omega}}{\partial t} + (\vec{u} \cdot \nabla)\vec{\omega} = (\vec{\omega}\cdot \nabla)\vec{u} + \nu\nabla^2\vec{\omega}
\end{equation}
This is also popularly wrriten as,
\begin{equation}
    \frac{D\vec{\omega}}{Dt} = (\vec{\omega}\cdot \nabla)\vec{u} + \nu\nabla^2\vec{\omega}
\end{equation}

\noindent \textbf{Non viscous 3D flows}: For non-viscous 3-D flows, we have the $\nu = 0$, then we the vorticity Barotropic equation as,
\begin{equation}
     \frac{D\vec{\omega}}{Dt} = (\vec{\omega}\cdot \nabla)\vec{u}
\end{equation}

\noindent \textbf{Non viscous 2D flows}: Consider the flow in 2D x-y plane: $\vec{u} = (u_x, u_y, 0))$. Then the vorticity $\vec{\omega} = \nabla \times \vec{u} = \left(\frac{\partial u_y}{\partial x} - \frac{\partial u_x}{\partial y}\right)\hat{z}$ is only in the $\hat{z}$ direction is thus perpendicular to the flow plane. Hence in this case, the vorticity can be thought of as a scalar. 
In this special case, we have that $(\vec{\omega}\cdot \nabla)\vec{u} = 0$. Thus the vorticity Barotropic equation in this case looks like,
\begin{equation}
    \frac{D\vec{\omega}}{Dt} = \frac{\partial \vec{\omega}}{\partial t} + (\vec{u} \cdot \nabla)\vec{\omega} = 0
\end{equation}


We now define the Barotropic equations for 2D flow in spherical coordinates. We divide the equations in stationary and in rotating sphere.

\subsection{Stationary sphere equations (in terms of vorticity)}

Let us now solely focus on the 2D non viscous incompressible flow in a stationary sphere. Consider the local flow on the sphere of radius $r$, $\vec{u} = (0, u_{\phi}, u_{\lambda}),$ where $u_{\phi}$ latitudinal velocity and $u_{\lambda}$ is the longitudinal velocity. The vorticity can be denoted as $\vec{\omega} = \zeta(\phi, \lambda, t)\hat{r}$ (where $\phi$ is the latitude with direction denoted by $\hat{\phi}$, $\lambda$ is the longitude with direction denoted by $\hat{\lambda}$ and $\hat{r}$ denotes that the vorticity is in the radial direction). The vorticity can then be expressed as $\vec{\omega} = \zeta(\phi, \lambda, t)\hat{r}$.

We now have to compute the term $\vec{u} \cdot \nabla \zeta$ in spherical coordinates. We can write the gradient of $\zeta$ in geographical coordinates as,
\begin{equation}
    \nabla \zeta = \frac{1}{r}\frac{\partial \zeta}{\partial \phi}\hat{\phi} + \frac{1}{r\cos\phi}\frac{\partial \zeta}{\partial \lambda} \hat{\lambda}
\end{equation}

Now the term $\vec{u} \cdot \nabla \zeta = \frac{u_{\phi}}{r}\frac{\partial \zeta}{\partial \phi} + \frac{u_{\lambda}}{r\cos\phi}\frac{\partial \zeta}{\partial \lambda}$. Thus the Barotropic equation in spherical coordinates are,
\begin{equation}
    \frac{\partial \zeta}{\partial t} + \frac{u_{\phi}}{r}\frac{\partial \zeta}{\partial \phi} + \frac{u_{\lambda}}{r\cos\phi}\frac{\partial \zeta}{\partial \lambda} = 0
    \label{Eq:baro2dsphere}
\end{equation}

Next, lets represent this function with respect to the streamline function. The streamline function $\psi$ is related to the vorticity $\zeta = \nabla^2 \psi$. In two dimensions, the vorticity is only in the radial direction, hence in geographical coordinates, it can be written as,
\begin{equation}
    \zeta = \nabla \times \vec{u} = \frac{1}{r \cos \phi}\left(\frac{\partial u_\phi}{\partial \lambda} - \frac{\partial}{\partial \phi}( u_\lambda \cos\phi)\right)
\end{equation}
Now, representing $\nabla^2 \psi$ in spherical coordinates, we have,
\begin{equation}
    \nabla^2 \psi = \frac{1}{r^2 \cos\phi} \frac{\partial}{\partial \phi}\left(\cos\phi \frac{\partial \psi}{\partial \phi}\right) + \frac{1}{r^2\cos^2\phi}\frac{\partial^2 \psi}{\partial \lambda^2}
\end{equation}
From the equality $\zeta = \nabla^2 \psi$, we have that,
\begin{equation}
    u_\phi = \frac{1}{r\cos\phi}\frac{\partial \psi}{\partial \lambda}
\end{equation}
and,
\begin{equation}
    u_{\lambda} = -\frac{1}{r}\frac{\partial \psi}{\partial \phi}
\end{equation}
Now we can represent the Eq.~\eqref{Eq:baro2dsphere} in terms of stream function $\psi$ and $\zeta$ as,
\begin{equation}
    \begin{split}
        \frac{D\zeta}{Dt} &= \frac{\partial \zeta}{\partial t} + \frac{u_{\phi}}{r}\frac{\partial \zeta}{\partial \phi} + \frac{u_{\lambda}}{r\cos\phi}\frac{\partial \zeta}{\partial \lambda} \\
        &= \frac{\partial \zeta}{\partial t} + \frac{1}{r^2\cos\phi}\left(\frac{\partial \psi}{\partial \lambda}\frac{\partial \zeta}{\partial \phi} - \frac{\partial \psi}{\partial \phi}\frac{\partial \zeta}{\partial \lambda} \right) \\
        &= \frac{\partial \zeta}{\partial t} + J(\psi, \zeta)
    \end{split}
\end{equation}
where $J(\psi, \zeta)$ is the Jacobian which is defined as,
\begin{equation}
    J(a, b) = \frac{1}{r^2 \cos\phi}\left(\frac{\partial a}{\partial \lambda}\frac{\partial b}{\partial \phi} - \frac{\partial a}{\partial \phi}\frac{\partial b}{\partial \lambda} \right)
\end{equation}
Now, we can write the final barotropic equation in terms of the Jacobian as follows,
\begin{equation}
    \frac{\partial \zeta}{\partial t} = -J(\psi, \zeta)
    \label{Eq:stationarybaro}
\end{equation}

\subsection{Stationary sphere equations (in terms of streamline)}

The above Barotropic equation in terms is expressed in terms of vorticity and streamline. However, given the extra equality condition $\zeta = \nabla^2 \psi$, we can express the above equation completely in terms of the streamline function alone. Let us rewrite the $\nabla^2 \psi$ in the spherical coordinates,
\begin{equation}
\begin{split}
    \nabla^2 \psi &= \frac{1}{r^2 \cos\phi} \frac{\partial}{\partial \phi}\left(\cos\phi \frac{\partial \psi}{\partial \phi}\right) + \frac{1}{r^2\cos^2\phi}\frac{\partial^2 \psi}{\partial \lambda^2} \\
    &= -\frac{1}{r^2}\tan\phi\frac{\partial \psi}{\partial \phi} + \frac{1}{r^2}\frac{\partial^2 \psi}{\partial \phi^2} + \frac{1}{r^2\cos^2\phi}\frac{\partial^2 \psi}{\partial \lambda^2} 
\end{split}
\end{equation}
Now let us rewrite the \gls{lhs} of Eq.~\eqref{Eq:stationarybaro},
\begin{equation}
\begin{split}
     \frac{\partial \zeta}{\partial t} &= \frac{\partial \nabla^2 \psi}{\partial t} \\
     &= -\frac{1}{r^2}\tan\phi\frac{\partial^2 \psi}{\partial \phi \partial t} + \frac{1}{r^2}\frac{\partial^3 \psi}{\partial \phi^2 \partial t} + \frac{1}{r^2\cos^2\phi}\frac{\partial^3 \psi}{\partial \lambda^2 \partial t}
\end{split}
\end{equation}

The RHS has the Jacobian term $J(\psi, \zeta)$ which has the terms, $\frac{\partial \zeta}{\partial \phi}$ and $\frac{\partial \zeta}{\partial \lambda}$. Let us calculate these terms individually,
\begin{equation}
\begin{split}
       \frac{\partial \zeta}{\partial \phi} &= \frac{\partial \nabla^2 \psi}{\partial \phi} \\
       &= -\frac{1}{r^2 \cos^2\phi} \frac{\partial \psi}{\partial \phi} - \frac{1}{r^2}\tan\phi\frac{\partial^2 \psi}{\partial \phi^2} + \frac{1}{r^2}\frac{\partial^3 \psi}{\partial \phi^3} \\ &+ \frac{2\tan\phi}{r^2 \cos^2\phi}\frac{\partial^2 \psi}{\partial \lambda^2}  + \frac{1}{r^2\cos^2\phi}\frac{\partial^3 \psi}{\partial \lambda^2 \partial \phi} 
\end{split}
\end{equation}
Similarly, the term $\frac{\partial \zeta}{\partial \lambda}$ is the following,
\begin{equation}
    \begin{split}
         \frac{\partial \zeta}{\partial \lambda} &= \frac{\partial \nabla^2 \psi}{\partial \lambda} \\
         &= -\frac{1}{r^2}\tan\phi\frac{\partial^2 \psi}{\partial \phi \partial \lambda} + \frac{1}{r^2}\frac{\partial^3 \psi}{\partial \phi^2 \partial \lambda} + \frac{1}{r^2\cos^2\phi}\frac{\partial^3 \psi}{\partial \lambda^3} 
    \end{split}
\end{equation}
Combining all of these in the Eq.~\eqref{Eq:stationarybaro}, we obtain,
\begin{equation}
\small
    \begin{split}
      &  -\tan\phi\frac{\partial^2 \psi}{\partial \phi \partial t} + \frac{\partial^3 \psi}{\partial \phi^2 \partial t} + \frac{1}{\cos^2\phi}\frac{\partial^3 \psi}{\partial \lambda^2 \partial t}  \\
      &+ \frac{1}{r^2 \cos\phi}\frac{\partial \psi}{\partial \lambda}\biggl( -\frac{1}{\cos^2\phi} \frac{\partial \psi}{\partial \phi} - \tan\phi\frac{\partial^2 \psi}{\partial \phi^2} + \frac{\partial^3 \psi}{\partial \phi^3} \\ &+ 2\tan\phi\frac{1}{\cos^2\phi}\frac{\partial^2 \psi}{\partial \lambda^2}  + \frac{1}{\cos^2\phi}\frac{\partial^3 \psi}{\partial \lambda^2 \partial \phi}\biggr) \\
      & - \frac{1}{r^2 \cos\phi}\frac{\partial \psi}{\partial \phi}\left(-\tan\phi\frac{\partial^2 \psi}{\partial \phi \partial \lambda} + \frac{\partial^3 \psi}{\partial \phi^2 \partial \lambda} + \frac{1}{\cos^2\phi}\frac{\partial^3 \psi}{\partial \lambda^3} \right) = 0
    \end{split}
\label{Eq:stationarybarostreamline}
\end{equation}

\subsection{Rotating sphere equation (in terms of vorticity)}

The above equation was in the stationary sphere assumption. But in the realistic setting, one needs to consider the motion of the earth surface. This induces an extra vorticity term which is due to the angular motion of the earth. The total vorticity is then written as $\zeta_{\textsf{tot}} = 2\vec{\Omega}\cdot\hat{r} + \nabla^2\psi = 2\Omega\sin\phi + \zeta$. Here refer $f = 2\Omega\sin\phi$. Thus the barotropic equation by taking the absolute vorticity term $\zeta_{\textsf{tot}}$ is:
\begin{equation}
    \begin{split}
        \frac{\partial \zeta_{\textsf{tot}}}{\partial t} &= \frac{\partial \zeta}{\partial t} = -J(\psi, \zeta_{\textsf{tot}}) = \\
        & = -J(\psi, f + \zeta) = -\beta u_{\phi} - J(\psi, \zeta)
    \end{split}
    \label{Eq:rotationbaro}
\end{equation}
where $\beta = \frac{2\Omega\cos\phi}{r}$.

Note that, the rotating sphere equation can just be written in terms of $\psi$ and $\zeta$,
\begin{equation}
    \frac{\partial \zeta}{\partial t} = -\frac{2\Omega}{r^2}\frac{\partial \psi}{\partial \lambda} - J(\psi, \zeta)
\end{equation}

\subsection{Rotating sphere equation (in terms of streamline)}

From the derivation done in Eq.~\eqref{Eq:stationarybarostreamline}, we rewrite the Eq.~\eqref{Eq:rotationbaro} in terms of the streamline and recover Eq.~\eqref{eqn:bve_as_streamline}
\begin{equation}
\small
       \begin{split}
      &  -\tan\phi\frac{\partial^2 \psi}{\partial \phi \partial t} + \frac{\partial^3 \psi}{\partial \phi^2 \partial t} + \frac{1}{\cos^2\phi}\frac{\partial^3 \psi}{\partial \lambda^2 \partial t} + 2\Omega \frac{\partial \psi}{\partial \lambda}  \\
      &+ \frac{1}{r^2 \cos\phi}\frac{\partial \psi}{\partial \lambda}\left( -\frac{1}{\cos^2\phi} \frac{\partial \psi}{\partial \phi} - \tan\phi\frac{\partial^2 \psi}{\partial \phi^2} + \frac{\partial^3 \psi}{\partial \phi^3}\right. \\
      & \left. + 2\tan\phi\frac{1}{\cos^2\phi}\frac{\partial^2 \psi}{\partial \lambda^2}  + \frac{1}{\cos^2\phi}\frac{\partial^3 \psi}{\partial \lambda^2 \partial \phi} \right) \\
      & - \frac{1}{r^2 \cos\phi}\frac{\partial \psi}{\partial \phi}\left(-\tan\phi\frac{\partial^2 \psi}{\partial \phi \partial \lambda} + \frac{\partial^3 \psi}{\partial \phi^2 \partial \lambda} + \frac{1}{\cos^2\phi}\frac{\partial^3 \psi}{\partial \lambda^3} \right) \\
      &= 0.
    \end{split}
\label{Eq:rotationbarostreamline}
\end{equation}

\section{Data sets}\label{app:data_sets}

\subsection{Generation of real global weather data evolving under the \gls{bve}}\label{app:real_world_data}

In this work our underlying equation is the \gls{bve}, which is only an approximation of the true nature of atmospheric physics. Therefore, even a perfect \gls{ml} solution would not exactly represent the real world future times. To account for this, we generate a reference data set by running our \gls{sem} solver (see Appendix \ref{app:spectral}) on a real-world initial vorticity state, producing dynamics that evolve under the \gls{bve}.

The initial state is the relative vorticity of the ERA5 hourly relative vorticity on 15th July 1980~\cite{hersbach2018era5} at a pressure level of 50 kPa. Without any processing, this data set contains 1148 longitudinal ($\lambda$) $\times$ 574 latitudinal ($\phi$) $\times$ 24 temporal ($t$) points for a total of $>10^7$ data points. Since the \gls{ml} training time for SL scales linearly with the number of data points, achieving reasonable run times requires us to downsample the spatial dimensions of the data. Such downsampling is achieved using the scikit-image \texttt{block\_reduce} function, for which square blocks of points are replaced by a single point which is the mean of the vorticities in that block. The length of the block is called the downsampling factor, since it is the factor by which the number of points in each dimension is reduced, and the total number of points are reduced by the square of the downsampling factor. 

Choosing the correct downsampling factor is important, as fewer data points leads to quicker training times. However, if downsampling too far, the important low-level features of the real-world vorticity are removed. This not only makes the data under study less interesting, but the \gls{sem} evolution of only high-level features leads to a trivial solution of the \gls{bve} where the vorticity is smoothed into one constant value across the globe. We find through trial and error that the largest downsampling factor which can be used, whilst allowing the \gls{sem} to propagate interesting low-level feature, is 13. This results in a data set of 89 ($\lambda$) $\times$ 45 ($\phi$) grid points. The largest of any of the boxes formed by this grid will be the longitude at the equator, corresponding to a size of $\frac{C_E}{89} = 450$km where $C_E = 40,075$km is the circumference of Earth at the equator. This means that our model has a minimum spatial resolution of $(450\text{km} \times 360^\circ) / C_E = 4.04^\circ$. We propagate this lower-resolution initial state using \gls{sem} time steps of 400s (the \gls{sem} solver uses units of seconds), taking snapshots every 3600s (1 hour), for a total of 23h of evolution. This leads to a data set we refer to as the \gls{rwd}, a set of $89 \times 45 \times 24=96{,}120$ points that represent the \gls{bve} evolution from a real-world initial state.

\subsection{Generation of artificial data evolving under the \gls{bve}}\label{app:artificial_data}

Whilst the ultimate goal of any weather modelling is to be applied to the physical world, solving the BVE from a real initial state is at the moment infeasible with current overheads of quantum computing and simulating quantum computers. Instead, in this work we consider an artificial initial global weather state, which we would like to have both local and global dynamics. To generate this state, we first take inspiration from the known single-mode analytic solution to the BVE, which on the unit sphere has the form
\begin{equation}\label{eqn:general_analytic_soln}
    \psi = P^m_l (\sin \phi) \cdot (\cos m\lambda \cos \sigma t + \sin m\lambda \sin \sigma t),
\end{equation}
where $P^m_l$ corresponds to the associated Legendre polynomial with modes $m,l$ and $\sigma$ is given by the dispersion formula
\begin{equation}
    \sigma = \sigma^m_l = -\frac{2 \Omega m}{l(l+1)}.
\end{equation}

Whilst this solution can be used to validate our \gls{dqc} solver, it in itself represents a fairly uninteresting evolution to solve for, since it contains only trivial global dynamics. In order to generate a richer (artificial) data set, we create an initial stream function state by combining multiple modes of the analytic solution. Specifically, first we construct the analytic solution generate the stream function $\psi$ at time $t=0$ over the sphere:
\begin{equation}
    \psi(t=0) = \sum_{m,l} P^m_l (\sin \phi) \cdot (\cos m\lambda),
\end{equation}
summing the $m=l=1$ and $m=1$, $l=2$ modes. 
This initial state is defined by a grid of 100 latitude ($\phi$) points and 200 longitude ($\lambda$) points over the entire sphere $\lambda \in \{0, 360\}, \phi \in \{-90, 90\}$.
Note that such summations of the single-mode analytic solutions do not satisfy the \gls{bve}, it is simply a convenient way to generate more interesting artificial data. 

Next, we translate the stream function values to vorticity using the relationship:
\begin{equation}\label{eqn:zeta_wrt_psi}
\begin{split}
    \zeta &= \frac{1}{r^2 \cos\phi} \frac{\partial}{\partial \phi}\left(\cos\phi \frac{\partial \psi}{\partial \phi}\right) + \frac{1}{r^2\cos^2\phi}\frac{\partial^2 \psi}{\partial \lambda^2} \\
    &= -\frac{1}{r^2}\tan\phi\frac{\partial \psi}{\partial \phi} + \frac{1}{r^2}\frac{\partial^2 \psi}{\partial \phi^2} + \frac{1}{r^2\cos^2\phi}\frac{\partial^2 \psi}{\partial \lambda^2}.
\end{split}
\end{equation}
Finally, this initial vorticity is fed into the \gls{sem} code along with various \gls{sem} parameters including the spectral truncation and the size of the evolution time step $\Delta t = 0.001$ (see Appendix \ref{app:spectral} for details).
Since we are working with an artificial system, the \gls{sem} solver evolves in unitless time steps. We also assuming a unit sphere $r=1$ and redefine the global rotation to be $\Omega=1$. 
The value of $\psi$ and $\zeta$ is then recorded at time intervals of 0.3, up to a maximum time $t_m = 3$. Finally, this data set is then spatially down-sampled to a smaller grid of $14 \times 25$ points to reduce the \gls{ml} training overhead. This leaves a total of 14 ($\phi$) $\times$ 25 ($\lambda$) $\times$  11 ($t$) grid points.

Overall, the SEM evolution produces dynamics with both global dynamics and local swirling patterns of two pairs of positive-negative stream function poles that rotate around one another (e.g. see Fig.~\ref{fig:dqc_result}).



\section{Spectral methods}
\label{app:spectral}

The dynamics of non-divergent flows on a rotating sphere are described by the conservation of absolute vorticity given in the \gls{bve}. One approach to solving this is through an expansion of the vorticity as a sum of spherical harmonics
\begin{equation}
\begin{split}
    \zeta(\phi, \lambda, t) &= \sum_{m=-L}^{L}\sum_{n=|m|}^{L} \zeta^m_n(t)Y^m_n(\lambda, \phi) \\
 &= \sum_{m=-L}^{L}\sum_{n = |m|}^{L} \zeta^m_n(t) P^m_n(\sin\phi)e^{im\lambda},
    \label{Eq:spherical_harmonics}
    \end{split}
\end{equation}
where $m$ is the order or the azimuthal wavenumber, $n$ is the degree and $L$ is the largest degree {included in the truncated expansion, under the usual expectation that a higher $L$ can lead to a more accurate reproduction of $\zeta$ at the cost of a more complex model}.  $Y^m_n$ is the eigenfunction of the Laplacian on the unit sphere with eigenvalue $-n(n+1)$. The $P^m_n$ are the associated Legendre polynomials. Further still, we can use the condition that $\zeta$ is real to enforce $\zeta^{-m}_n = \zeta^{m}_n$. Thus the above Eq.~\eqref{Eq:spherical_harmonics} can be rewritten as
\begin{equation}
\begin{split}
    \zeta(\phi, \lambda, t) &= \sum_{n=0}^{L}\zeta_{n}^0 P_{n}^0(\sin \phi) \\ &+2\sum_{m=1}^{L}\sum_{n=m}^{L}P_{n}^m(\sin\phi)\Re(\zeta_{n}^m(t)e^{im\lambda}).
    \label{Eq:spherical_harmonics 2}
    \end{split}
\end{equation}
This spectral representation can be used to solve the \gls{bve} through a \gls{sem}. In the SEM, the spatial domain is discretized into elements, each represented by one of the basis functions. Spatial derivatives can be analytically computed within each element, which leads to a semi-discrete system of ordinary differential equations (ODEs), which may be integrated in time using time-stepping schemes. Our SEM benchmark is based on publicly available github code \footnote{ \url{https://github.com/jweyn/DLWP-CS/tree/master/DLWP/barotropic}
}, which works as follows:
\begin{enumerate}
    \item Input the initial $t=0$ vorticity of the system to be evolved.
    \item Define the grid of $t$, $\phi$, $\lambda$ coordinates, the total evolution time, the step size of each evolution $\Delta t$ and the number of basis functions to use.
    \item Convert the grid vorticity into the spectral vorticity.
    \item Evolve the spectral vorticity a step forward by solving the discretised version of Eq.~\eqref{Eq:rotationbarostreamline}. This step forward uses the triangular truncation technique and Robert coefficient smoothing to prevent the resulting spectral vorticity errors from accumulating~\cite{ecmwf_numerical}.
    \item Convert the stepped-forward spectral vorticity back to a grid vorticity. 
    \item Repeat steps 3-5 until we the total evolution time is reached. This final grid vorticity is then returned as the evolved vorticity. 
\end{enumerate}

We note that the time step parameter $\Delta t$ is perhaps the most important for the \gls{sem} and picking it correctly is a balancing act: too large and numerical instability will cause the vorticity to explode, too small and the computation will take too long to run. To find an optimal solution, we test increasingly large $\Delta t$ values on a trial evolution of only 100 time steps and observe the largest value that an instability does not occur. The specific values found for each data set are given in Appendix \ref{app:real_world_data} and \ref{app:artificial_data} respectively.

\section{Definition of Figures of Merit}\label{app:fom}

We establish two \gls{fom}, both based on the difference between a predicted metric (stream function or vorticity) and the reference data. The first is the mean relative error (MRE) compared to the \gls{sem} median. This is defined for a single time as
\begin{equation}\label{eq:MRE}
    \text{MRE} = \frac{1}{N}\sum^N \left| \frac{\psi_{NN} - \psi_{SEM}}{\tilde{|\psi|}_{SEM}}\right|,
\end{equation}
where $\psi_{NN}$ is the stream function predicted by the trained \gls{pinn} model, ${|\tilde{\psi}|_{SEM}}$ is the median of the absolute stream function predicted by the \gls{sem} model and the summation $N$ is over the spatial grid points. Note that whilst our equations are written in terms of stream function, the same \gls{fom} are defined for the vorticity.

We also define the Pearson product-moment correlation coefficient (PPMCC) as
\begin{equation}
    R_{ij} = \frac{C_{ij}}{\sqrt{C_{ii}C_{jj}}},
\end{equation}
where $C_{ij}$ is the covariance between random variables $i$ and $j$. In the context of this work, we define the stream function of the \gls{qnn} and \gls{sem} solutions for a single grid point as random variables and consider the values at each time step as samples of the variables. Thus, for each grid point we can calculate the PPMCC, computing how correlated the two solutions are over time. Since the PPMCC is normalised, a value of $-1$, 0 and 1 represents perfect anti-correlation, no-correlation and perfect correlation respectively. Numerically we find that models with high PPMCC correspond to models that also score low in MRE. This agreement between our correlation-based \gls{fom} and distance-based \gls{fom} gives us a powerful toolkit to analyse the quality of the machine learning models trained.

\end{document}

%% file: glossary.tex
\newacronym{ifs}{IFS}{Integrated Forecasting System}
\newacronym{ecmwf}{ECMWF}{European Centre for Medium-Range Weather Forecasts}

\newacronym{bve}{BVE}{barotropic vorticity equation}
\newacronym{sem}{SEM}{spectral element method} 

\newacronym{nn}{NN}{neural network}
\newacronym{pinn}{PINN}{physics-informed neural network}
\newacronym{fm}{FM}{feature map}
\newacronym{ml}{ML}{machine learning}
\newacronym{hea}{HEA}{Hardware Efficient Ansatz}
\newacronym{sciml}{SciML}{scientific machine learning}
\newacronym{ufa}{UFA}{universal function approximator}

\newacronym{qml}{QML}{quantum machine learning}
\newacronym{dqc}{DQC}{differential quantum circuits}
\newacronym{pqc}{PQC}{parameterised quantum circuit}
\newacronym{qcl}{QCL}{quantum circuit learning}
\newacronym{nisq}{NISQ}{noisy intermediate scale quantum}
\newacronym{qnn}{QNN}{quantum neural network}
\newacronym{qsciml}{QSciML}{quantum scientific machine learning}

\newacronym{rwd}{RWD}{real-world data}
\newacronym{fom}{FOM}{figures of merit}
\newacronym{pde}{PDE}{partial differential equation}
\newacronym{lhs}{LHS}{left-hand side}
\newacronym{tffm}{TFFM}{trainable-frequency feature map}
\newacronym{mse}{MSE}{mean squared error}